\documentclass[doublecol]{epl2} 

\usepackage[english]{babel}

\title{Entanglement of electron pairs extracted from a many-body \\system}

\author{A. Ram{\v s}ak\inst{1,2} \and J. Mravlje\inst{2} \and T.
Rejec\inst{1,2} \and A. Lautar\inst{1}}

\shortauthor{A. Ram{\v s}ak \etal}

\institute{                    
  \inst{1} Faculty of Mathematics and Physics, University of
  Ljubljana, Ljubljana, Slovenia\\
  \inst{2} Jo{\v z}ef Stefan Institute, Ljubljana, Slovenia
}

 
\abstract{Entanglement of spins is analyzed for two electrons
extracted from a mixed many electron state by projecting onto the
two-electron subspace. The concurrence formulae are expressed in a
compact form for states with a well defined square of the total spin
projection. As an example, the thermal entanglement for a qubit pair
with an anisotropic Heisenberg and the Dzyaloshinskii-Moriya
interactions in an inhomogeneous magnetic field is given analytically. 
Remarkably, the concurrence of a pair of electrons with antiparallel spins and in a
delocalised orbital state is given by the scalar product of the state
with its spin-flipped state and not with the time-reversed state.}

\begin{document}

\maketitle

Quantum entanglement is considered a key resource for quantum
cryptography and quantum computation \cite{nielsen}. Quantifying
entanglement \cite{plenio06,amico08} and identifying maximally
entangled states is thus important in the planning of devices. On the
other hand, the research of the entanglement and its decoherence can
{also} lead  to a better understanding of the foundations of physics, for
example of the quantum to classical crossover \cite{zurek03} and of
the origins of the thermodynamic laws \cite{brandao08}.

Quantum dot arrays, controlled by electrical gating, are,
due to their scalability, promising candidates for operational devices
\cite{loss98,amasha08}. The building blocks are coupled few electron
quantum dots enabling full control over individual electrons
\cite{hanson07,nowack}. In a quantum dot the qubit is usually
represented by the spin of an electron. Alternatively, charge
pseudo-spin entanglement in double quantum dots can also be exploited
\cite{mravljedqd}, but is prone to the decoherence due to the Coulomb
interaction with the environment.

The description of electrons by the spin degrees of freedom only is a
simplification valid when the electrons are localized and the charge
fluctuations are negligible. In general, both orbital and spin degrees
of freedom are present, but if one is interested in the spin
entanglement only, one should trace out the spatial dependence.
Typical examples are a recently proposed route to generation of
perfectly entangled electron pairs by the use of acoustic waves in the
surface of a GaAs/AlGaAs structure \cite{giavaras06} or elastic
scattering of electrons in semiconducting carbon nanotube structures
with orbital degeneracy \cite{habgood08}. For a special case, where
there are precisely two electrons in a pure state on the lattice, the
entanglement can be given by simple formulae expressed in terms of the
wave-function \cite{ramsak06}. Such formulae can be applied to the
determination and optimization of entanglement generation between
static and flying qubits in one dimensional systems\cite{jrr05,buscemi07}. 
They also enable the analysis of the entanglement between qubit pairs in 
various double quantum dot configurations coupled to external leads \cite{2d}.

Here we are interested in the entanglement of an electron pair
extracted from a many-body state, which can be, for example, an open
system of interacting electrons in a solid state structure of several
coupled quantum dots. In particular, we take that the measurement
apparatus extracts precisely one electron from each of two
non-overlapping regions of the structure -- domains $A$ and $B$. The
state of the system is arbitrary and includes fluctuations of
electrons between the domains or between the domains and the
environment, which introduces spin and charge fluctuations to the
subsystem $A \cup B$.

{In this letter we express the reduced density matrix of
two spin-qubits in terms of
projected spin-spin correlators which allows
the analysis of entanglement of qubit pairs extracted from a general many electron state.
The corresponding concurrence is then given explicitly for systems conserving the square of the 
total spin projection, which is illustrated by several examples.}

First consider two separated electrons, one from domain $A$ and the
other from domain $B$, with spin states labeled by $s=\pm {1\over2}$
and $t=\pm {1\over2}$, respectively. Let the electrons be in a pure
state expressed in the standard basis
$|\mu\rangle\equiv|st\rangle\in\left\{\left|\uparrow\uparrow\right\rangle,
\left|\uparrow\downarrow\right\rangle,
\left|\downarrow\uparrow\right\rangle,
\left|\downarrow\downarrow\right\rangle\right\}$ for $\mu\in\{1,2,3,4\}$,
as
\begin{equation}
|\psi\rangle=\sum_\mu \alpha_\mu |\mu\rangle. \label{purepsi}
\end{equation}
Because the electrons are in a state completely determined by the spin
degrees of freedom only, the entanglement can be quantified with the
entanglement of formation or, equivalently, with the concurrence
$C=2|\alpha_1\alpha_4-\alpha_2\alpha_3|$ \cite{bennett96}.  In
general, two spins may be a subsystem of a larger system with many
degrees of freedom and the subsystem is then described by a reduced
$4\times4$ density matrix $\rho$. In this case the concurrence is
given by the Wootters formula \cite{wootters98},
\begin{equation}
C=\max(0,2 {\lambda}_{max}-\sum_{j=1}^4 {\lambda}_j), \label{cmax}
\end{equation} 
where $\lambda_j$ are the square roots of the eigenvalues of the
non-Hermitian matrix $\rho\tilde{\rho}$ among which $\lambda_{max}$ is
the largest, and $\tilde{\rho}$ is the time-reversed density matrix
$\rho$.

In this paper we consider the electrons in domains $A$ and $B$ as a
subsystem of a total system described by a density matrix
\begin{equation}
\rho_{tot}=\sum_n p_n |n\rangle\langle n|.\nonumber \label{rotot}
\end{equation} 
Then a projective measurement is performed by an apparatus which
extracts an electron pair: one electron from $A$ and another one from
$B$ in such a way that after the projection, the system is in one of
the normalized states ${\cal P} |n\rangle/\sqrt{\langle n|{\cal P}
|n\rangle}$, where the projector ${\cal P}$ projects onto the subspace
where in each of the domains $A$ and $B$ there is exactly one
electron. What we are left with is the density matrix
\begin{equation}
  \rho_{\cal P}=\sum_n q_n{ {\cal P}|n\rangle\langle n|{\cal P} \over
  \langle n|{\cal P} |n\rangle}= {{\cal P}\rho_{tot}{\cal P} \over
  {\textrm{Tr}}{\cal P} \rho_{tot}},\nonumber\label{rop}
\end{equation}
where $q_n=P(n|{\cal P})$ is the conditional probability that after
the measurement the system will be in the projected state ${\cal
P}|n\rangle$, i.e., that the electrons were extracted from the state
$|n\rangle$. This probability is given by the Bayes' formula
\cite{nielsen}, $P(n|{\cal P}) =P({\cal P}|n) p_n/\langle {\cal P}\rangle$, where $P({\cal
P}|n)= \langle n|{\cal P} |n\rangle$ is the conditional probability
for a single occupancy of each of the domains for a particular state
$|n\rangle$ and $\langle {\cal P}\rangle=\sum_n p_n \langle n|{\cal P}
|n\rangle={\textrm{Tr}}{\cal P} \rho_{tot}$ is the probability that
the apparatus will click, i.e., that the desired two electrons will be
extracted.

The projected states ${\cal P}|n\rangle$ read
\begin{equation}
{\cal P}|n\rangle=\sum_{ij,st}\psi_{n,ij}^{st}c_{i s}^{\dagger}c_{j
t}^{\dagger}|\Phi_{n,ij}^{st}\rangle, \nonumber\label{pn}
\end{equation}
where  {the operators
$c_{ is}^{\dagger}$ and $c_{
jt}^{\dagger}$ create electrons with spin $s$ at sites $i\in A$ and
with spin $t$ at sites $j\in B$, respectively, and, being
ordinary electron creation operators, obey the fermionic rule  
$c_{ is}c_{i's'}^{\dagger}+c_{i's'}^{\dagger}c_{ is}=\delta_{ii'}\delta_{ss'}$.}

The number of sites
within the domains is arbitrary. {States 
$|\Phi_{n,ij}^{st}\rangle$ are normalized and represent empty domains $A$ and $B$ with the rest of the system in
an arbitrary configuration. In general, these vacuum states may be different for each of the states $|n\rangle$ and also for
each particular occupation of pairs of sites $(i,j)$ within the domains.  The projector ${\cal P}$ removes from $|n\rangle$
all components except those where each of the domains is occupied by precisely one electron and may be written explicitly as
\begin{equation}
{\cal P}=\prod_{k=0,k\ne1}^{N_A}
\frac{k-\hat{n}_A}{k-1}\prod_{k=0,k\ne1}^{N_B}\frac{k-\hat{n}_B}{k-1},
\nonumber\label{pp}
\end{equation}
where $\hat{n}_{A(B)}=\sum_{l\in A(B),s} c_{ls}^{\dagger}c_{ls}$ is the number operator for domains $A(B)$ and
$N_{A(B)}$ is the maximum possible number of electrons in $A(B)$}.

Being interested in the spin entanglement we consider the reduced
density matrix where only the spin degrees of freedom are retained,
$\rho=\sum_{\mu\nu} \rho_{\mu\nu} |\mu\rangle \langle \nu |$, with
$\rho_{\mu \nu}\equiv\rho_{(s t)(s' t')}$ and
\begin{equation}
\rho_{(s t)(s' t')} = {1\over \langle {\cal P}\rangle} \sum_{n,ij} p_n \langle \Phi_{n,ij}^{s't'}|\Phi_{n,ij}^{st}\rangle
(\psi_{n,ij}^{s' t'})^*\psi_{n,ij}^{s t}.  \nonumber\label{reduced}
\end{equation}
This formula is useful if the wave functions are known. However, in
some cases it is possible to determine various correlation functions
of the system without an explicit knowledge of the wave functions.
Then it is advantageous to express the density matrix in terms of spin
correlators \cite{syljuasen},
\begin{eqnarray}
\rho={1\over \langle {\cal P}\rangle}
\left(
\begin{array}{llll} \langle P_{A}^{\uparrow}P_{B}^{\uparrow}\rangle & 
\langle P_{A}^{\uparrow}S_{B}^-\rangle &\langle S_{A}^- P_{B}^{\uparrow}\rangle  & \langle S_{A}^{-}S_{B}^{-}\rangle\\
\langle P_{A}^{\uparrow}S_{B}^+\rangle & \langle
P_{A}^{\uparrow}P_{B}^{\downarrow}\rangle&\langle
S_{A}^{-}S_{B}^{+}\rangle &\langle S_{A}^-P_{B}^{\downarrow}\rangle \\
\langle S_{A}^+ P_{B}^{\uparrow}\rangle & \langle
S_{A}^{+}S_{B}^{-}\rangle& \langle
P_{A}^{\downarrow}P_{B}^{\uparrow}\rangle &\langle
P_{A}^{\downarrow}S_{B}^-\rangle \\ \langle S_{A}^{+}S_{B}^{+}\rangle
&\langle S_{A}^+ P_{B}^{\downarrow}\rangle &\langle
P_{A}^{\downarrow}S_{B}^+\rangle & \langle
P_{A}^{\downarrow}P_{B}^{\downarrow}\rangle\\
\end{array}
\right),  \nonumber\label{romat}
\end{eqnarray}
where $\langle {\cal P}\rangle=\sum_{st}\langle P_A^{s}
P_B^{t}\rangle$ is the probability that in the subsystem $A\cup B$
there will be precisely two electrons, one in each of the domains. The
correlators are expressed as the expectation values of projected
operators in the sense $\langle {\cal O}\rangle\equiv\sum_n p_n
\langle n| {\cal P}{\cal O}{\cal P}|n\rangle$ where ${\cal O}$
consists of $A$-$B$ pairs of operators
\begin{eqnarray}
S_{A(B)}^{x,y,z}&=&\frac{1}{2}\sum_{l\in A(B),ss'}
\sigma_{ss'}^{x,y,z} c_{ls}^{\dagger}c_{ls'}, \nonumber\\ \label{sab}
P_{A(B)}^{s}&=&\sum_{l\in A(B)}\hat{n}_{l,s}(1-\hat{n}_{l,-s}). \nonumber
\end{eqnarray}
Here $\sigma_{ss'}^{x,y,z}$ are the Pauli matrices,
$S_{A(B)}^{\pm}=S_{A(B)}^x\pm i S_{A(B)}^y$ and {
$\hat{n}_{l,s}=c_{ls}^{\dagger}c_{ls}$} is the electron number operator.

The evaluation of the correlators is simplified if each of the domains
$A$ and $B$ consists of one site only, in which case ${\cal P}{\cal
O}{\cal P}={\cal O}$, i.e., the states corresponding to empty or
doubly occupied sites are projected away by the operator ${\cal O}$. A
prototype example is a Hubbard dimer, i.e., two electrons on two sites
and described by the Hubbard model, as studied by Zanardi
\cite{zanardi02}. Note, however, that the entanglement measures
introduced for fermionic systems where multiple occupancy is retained
\cite{schliemann01,zanardi02,gittings02,vedral03,naudts07} are
different from the entanglement of formation studied here.

The concurrence for the domains $A$ and $B$ is determined from the
projected density matrix by the Wootters formula, eq.~(\ref{cmax}).
In general, $\lambda_j$ can be computed numerically, but in some
cases, due to symmetry the density matrix simplifies and analytic
evaluation is possible.  Such symmetries were exploited in various
coupled spin systems on a lattice with translational and parity
invariance \cite{syljuasen03,osterloh02,roscilde04, amico04}.

In the present case of interacting electrons in coupled quantum dot
structures, translational and parity invariance is an exception. Still,
an analogous simplification is possible in a special case when the
density operator commutes with the square of the total spin projection
for $A\cup B$, $S^z=S_A^z+S_B^z$. For such {\it biaxial} systems
\cite{ziherl} with
\begin{equation}
[\rho,(S^z)^2]=0,  \nonumber\label{sz2}
\end{equation}
$\rho$ is a block matrix: $\rho_{12}=\rho_{13}=\rho_{24}=\rho_{34}=0$
or, equivalently, $\langle S_{A(B)}^{x,y}\rangle=0$ and $\langle
S_{A(B)}^{z}S_{B(A)}^{x,y}\rangle=0$.

The eigenvalues $\lambda_j^2$ of $\rho\tilde{\rho}$ (which, again, is
a block matrix) follow trivially from two decoupled blocks
corresponding to subspaces with parallel, $(S^z)^2 =1$ and $\{\mu=1,
\nu=4\}$, or antiparallel spins, $(S^z)^2=0$ and $\{\mu=2, \nu=3\}$.
The matrix elements $\rho_{\mu \nu}$ are interrelated \cite{landau},
$|\rho_{\mu \nu}|\le \sqrt{\rho_{\mu\mu} \rho_{\nu\nu}}$, which leads
to $\lambda_j=\sqrt{\rho_{\mu\mu}\rho_{\nu\nu}}\pm |\rho_{\mu\nu}|
$. The concurrence is then determined by
\begin{eqnarray}
C & = &
\max\left(0,C_{\uparrow\!\downarrow},C_{\parallel}\right)/\sum_{st}\langle P_A^{s}
P_B^{t}\rangle,\label{ccc}\\
C_{\uparrow\!\downarrow} & = & 2|\langle
S_{A}^{+}S_{B}^{-}\rangle|-2\sqrt{\langle
P_{A}^{\uparrow}P_{B}^{\uparrow}\rangle\langle
P_{A}^{\downarrow}P_{B}^{\downarrow}\rangle},\nonumber \\
C_{\parallel} & = & 2|\langle
S_{A}^{+}S_{B}^{+}\rangle|-2\sqrt{\langle
P_{A}^{\uparrow}P_{B}^{\downarrow}\rangle\langle
P_{A}^{\downarrow}P_{B}^{\uparrow}\rangle},\nonumber
\end{eqnarray}
which represents a generalisation of the result derived for the case
of precisely two delocalised electrons in a pure state
\cite{ramsak06}.

For {\it axially} symmetric systems, i.e., conserving the total spin
projection, $[\rho,S^z]=0$, the formula simplifies because $\langle
S_{A}^{+}S_{B}^{+}\rangle=0$ and
\begin{equation}
C = \max\left(0,C_{\uparrow\!\downarrow}\right)/\langle {\cal P}\rangle.\label{cc1}
\end{equation}
For SU(2) spin symmetric case, $\left[\rho,{\mathbf S}_{A}+{\mathbf
S}_{B}\right]=0$, the concurrence is completely determined by a single
spin invariant and $C=2\max\left(0,-\left\langle{\mathbf
S}_A\cdot{\mathbf S}_B\right\rangle/\langle {\cal P}\rangle-\frac{1}{4}\right)$.

In practice, several specific cases are of interest. Let us first
consider a special case of the total system being in a pure state
$\left|m\right\rangle$ containing only two electrons, i.e.,
$p_n=\delta_{nm}$ and $|\Phi_{m,ij}^{st}\rangle=|0\rangle$. We assume
the electrons are in a state with the amplitudes
$\psi_{m,ij}^{\uparrow\downarrow}=\alpha_2 \varphi_{ij}$,
$\psi_{m,ij}^{\downarrow\uparrow}=\alpha_3 \varphi_{ij}$ and
$\psi_{m,ij}^{\uparrow\uparrow}=\alpha_1 \chi_{ij}$,
$\psi_{m,ij}^{\downarrow\downarrow}=\alpha_4 \chi_{ij}$, where
$\varphi_{ij}$ and $\chi_{ij}$ are normalized. Then, if $\varphi_{ij}=
\chi_{ij}$, the concurrence is given by
$C=2|\alpha_1\alpha_4-\alpha_2\alpha_3|/\langle {\cal P}\rangle$ with $\langle {\cal P}\rangle=\sum_\mu
|\alpha_\mu|^2$, which is the pure spin-subsystem result, renormalized
due to the projection. If $\sum_{ij}\varphi_{ij}^\ast \chi_{ij}=0$ the
concurrence is $C=2
\left|\left|\alpha_1\alpha_4\right|-\left|\alpha_2\alpha_3\right|\right|/\langle {\cal P}\rangle$. Additionally,
if the state $\left|m\right\rangle$ is an eigenstate of $S^z$, the
concurrence simplifies further to $C=2|\alpha_2\alpha_3|/\langle {\cal P}\rangle$.

An interesting case in point is a pure state with a zero spin
projection, $S^z|m\rangle=0$, and $\langle{\cal P}\rangle=1$. Such
states are important, for example, in the realization of entangled
flying qubit pairs, when two initially unentangled electrons approach
each other and the interaction conserves $S^z$. The concurrence is
given by $2|\langle m|S_A^+S_B^-|m\rangle|$, but 
can also be expressed as 
\begin{equation}
C=\sqrt{\langle {\cal F}\rangle^2+{4}\langle m|{\bf
S}_A\times{\bf S}_B|m \rangle^2}. \label{fss}
\end{equation}
Here $\langle {\cal F}\rangle=\langle m|m_{\rm flip}\rangle$ is the scalar product
of the state
$|m\rangle$ with its spin-flipped state $|m_{\rm flip}\rangle={\cal
F}|m\rangle$ where the spin-flip operator ${\cal
F}=S_{A}^{-}S_{B}^{+}+S_{A}^{+}S_{B}^{-}$ reverses the spins in $A\cup
B$. 

For a special case $\langle {\bf S}_A\times{\bf S}_B \rangle=0$, the expression eq.~(\ref{fss})
resembles the result for a general pure spin-state,
eq.~(\ref{purepsi}), with the concurrence given by $C=|\langle
m|{\widetilde m}\rangle|$, where $|{\widetilde m} \rangle={\cal
T}|m\rangle$ is the time-reverse of $|m\rangle$ \cite{wootters98}.
The time reversal operator is given by $\exp[-i\pi(S_{A}^y+S_{
B}^y)]{\cal K}$, where ${\cal K}$ is the complex conjugation operator
\cite{sakurai}, which for the present case of two electrons with
antiparallel spins gives ${\cal T}=i{\cal F}{\cal K}$. For a special case
of a pure spin-state, where $\psi_{m,ij}^{\uparrow\downarrow} \propto
\psi_{m,ij}^{\downarrow\uparrow}$ and $\rho^2=\rho$, the conjugation
has no effect and the concurrence takes the customary form
$C=2|\alpha_2\alpha_3|=2|\alpha_2^\ast\alpha_3|$. In general, however,
if each of the domains consists of at least two sites, the overlap of
the state $|m\rangle$ with the spin-flipped state $| m_{\rm
flip}\rangle$ is different from its overlap with the time-reversed
state $|{\widetilde m}\rangle$ because the amplitudes in $|{\widetilde
m}\rangle$ are complex conjugated while in $|m_{\rm flip}\rangle$ they
are not. To be specific, let two electrons be in the state
\begin{equation}
|\Psi\rangle={1 \over 2}(c^\dagger_{1\uparrow} c^\dagger_{3\downarrow}+
c^\dagger_{1\downarrow} c^\dagger_{3\uparrow}+ic^\dagger_{2\uparrow} c^\dagger_{4\downarrow}+
ic^\dagger_{2\downarrow} c^\dagger_{4\uparrow})|0\rangle,\nonumber
\end{equation}
where the sites $1,2$ and ${3,4}$ represent the domains $A$ and $B$,
respectively. The correct expression for the concurrence is
$C=|\langle \Psi|\Psi_{\rm flip}\rangle|=1$ while the scalar product
with the time-reversed state is $\langle \Psi|{\widetilde
\Psi}\rangle=0$. Thus, in general the entanglement of electron pairs
is not related to the scalar product of a state with its time-reverse
but to the spin-flipped state only. This apparent disagreement with
the pure spin result is no paradox; it should simply be a warning and
a demonstration that the
electron pair with a general orbital extend can not be described by a
pure spin-state. 

Finally, we present an example where the biaxial symmetry $[\rho,
(S^z)^2]=0$ arises naturally in a system in thermal
equilibrium. Consider a pair of electrons in domains $A$ and $B$
representing a weakly coupled double quantum dot structure where
charge fluctuations between $A$ and $B$ and with the rest of the
system are negligible (i.e., $\langle {\cal P}\rangle=1$). Besides the
Coulomb interaction, the spin-orbit interaction may also be present
and a general effective Hamiltonian can contain the anisotropic
Heisenberg exchange interaction, the Dzyaloshinskii-Moriya term
\cite{moriya} and the coupling to an inhomogeneous external magnetic
field,
\begin{eqnarray}
H&=&J_x S^x_A S^x_B+J_y S^y_A S^y_B+J_z S^z_A S^z_B+\nonumber \\
\label{heis} & &+ {\bf B}_A \cdot {\bf S}_A+{\bf B}_B\cdot {\bf
S}_B+\\ & &+ {\bf D}\cdot ({\bf S}_A\times{\bf S}_B)+{\bf S}_A \cdot
\textsf{T}{\bf S}_B. \nonumber
\end{eqnarray}
We assume the magnetic field and ${\bf D}$ are parallel to the
$z$-axis and a symmetric tensor $\textsf{T}$ is of the form
\begin{eqnarray}
\textsf{T}=\left(
\begin{array}{lll} 0 & t & 0  \\ 
t & 0 & 0 \\
0 & 0 & 0\\
\end{array}
\right).\nonumber \label{calt}
\end{eqnarray}
In the basis $\{|\mu\rangle\}$ the Hamiltonian takes the form
\begin{eqnarray}
H=\left(
\begin{array}{llll} h_{11} & 0 & 0 & h_{14} \\ 
0 & h_{22} & h_{23} & 0\\
0 & h_{23}^* & h_{33} & 0\\
h_{14}^* & 0 & 0 & h_{44} \\
\end{array}
\right), \label{H}
\end{eqnarray}
which is the most general form of a Hamiltonian commuting with
$(S^z)^2$. The corresponding density matrix describing thermal
equilibrium is given by $\rho=\exp(-\beta H)/{\rm Tr}[\exp(-\beta
H)]$, where $\beta$ is the inverse temperature. The matrix elements of
the Hamiltonian eq.~(\ref{H}) are related to eq.~(\ref{heis}) as
$h_{11(44)}=J_z/4\pm(B_A+B_B)/2$, $h_{14}=(J_x-J_y)/4+i t/2$,
$h_{22(33)}=-J_z/4\pm(B_A-B_B)/2$, $h_{23}=(J_x+J_y)/4+i D/2$ with
${\bf B}_{A(B)}=B_{A(B)} \hat{\bf z}$ and ${\bf D}=D\hat{\bf z}$.

The problem decouples into two $2\times2$ subsystems and the resulting
two pairs of eigenenergies are given analytically,
\begin{eqnarray}
  \epsilon_{\{\mu\nu\}}&=&(h_{\mu\mu}+h_{\nu\nu})/2\pm x_{\{\mu\nu\}},\nonumber\\
  x_{\{\mu \nu\}}&=&[(h_{\mu \mu}-h_{\nu \nu})^2/4+|h_{\mu \nu}|^2]^{1/2}, \nonumber\label{xij}
\end{eqnarray}
where the subscript $\{\mu\nu\}$ denotes $\{14\}$ and $\{23\}$ for subspaces
with $(S^z)^2 =1$ and $(S^z)^2=0$, respectively.  With the
corresponding sets of eigenvectors, the density matrix elements are
easily expressed,
\begin{eqnarray}
|\rho_{\mu\nu}| & = &{|h_{\mu\nu}| \sinh \beta x_{\{\mu\nu\}} \over Z
  x_{\{\mu\nu\}}}e^{-\beta(h_{\mu\mu}+h_{\nu\nu})/2},
\nonumber\label{munu}\\ \rho_{\mu\mu}\rho_{\nu\nu} & =
&|\rho_{\mu\nu}|^2+Z^{-2}e^{-\beta(h_{\mu\mu}+h_{\nu\nu})},
\nonumber\label{mumu}
\end{eqnarray}
with
\begin{eqnarray}
Z&=&2e^{-\beta(h_{11}+h_{44})/2} \cosh \beta x_{\{14\}} +\nonumber\\
&+&2 e^{-\beta(h_{22}+h_{33})/2}\cosh \beta x_{\{23\}}. \nonumber
\end{eqnarray}
The concurrence is given by eq.~(\ref{ccc}),
$C=2\max(0,\left|\rho_{23}\right|-\sqrt{\rho_{11}\rho_{44}}, \left|\rho_{14}\right|-\sqrt{\rho_{22}\rho_{33}})$.

Some particular cases of this problem, i.e., an isotropic Heisenberg
model in a magnetic field and spin Hamiltonians with the
Dzyaloshinskii-Moriya interaction \cite{arnesen,wang, yang06}, were
analyzed before. For $J_x=J_y$ and $t=0$ the qubit pair is axially
symmetric and the concurrence simplifies to eq.~(\ref{cc1})
with 
\begin{equation}
C_{\uparrow\!\downarrow}= \left(|h_{23}|{{\sinh} \beta x_{\{23\}} \over x_{\{23\}}}
e^{-\beta J_z/4}-e^{\beta J_z/4}\right)/Z.
\end{equation}

In conclusion, we analyzed the spin entanglement of electron pairs
extracted from a system of electrons in a mixed state by projecting it
onto the two-electron subspace. The entanglement is quantified by the
concurrence obtained from the reduced density matrix which is
expressed in terms of projected spin-spin correlators for the
measurement domains and normalized by the probability that in each of
the two measurement domains there is precisely one electron. The
formalism is appropriate for the analysis of the entanglement of
formation for the domains of open fermionic systems allowing charge
fluctuations of the subsystems, as is, e.g., a system of coupled
quantum dots attached to external leads.

Simplified expressions are derived for systems with a good square of
spin projection.  The result for the most general case of the
corresponding two qubit system is given analytically, which
generalizes particular known cases. As an example, the thermal state
of a double quantum dot with anisotropic Heisenberg and
Dzyaloshinskii-Moriya interactions and in an inhomogeneous magnetic
field is considered. The concurrence is presented in a simple closed form.

Also considered is an electron pair in a pure, but orbitally
delocalized state. An example, relevant to the analysis of solid state
realizations of flying and static qubits, demonstrates that the
concurrence is given as the overlap of the state with its spin-flipped
state -- but not complex conjugated -- therefore not by its
time-reverse as is the case for pure spin-states.

\acknowledgments
We acknowledge the discussions with I. Sega and M. Nemev{\v s}ek and
the support from the Slovenian Research Agency under contracts Pl-0044
and J1-0747.


\begin{thebibliography}{10}
\bibitem{syljuasen} 
  The application of correlators to the present case of
  indistinguishable fermions is a straightforward generalization of the
  approach applied in the case of spin systems \cite{syljuasen03}.
\end{thebibliography}
\end{document}